\documentclass[aps,prl,twocolumn]{revtex4}

\usepackage{graphicx}

\usepackage{epsf}

\newcommand{\re}[1]{(\ref{#1})}

\newcommand{\up}{\uparrow}

\newcommand{\dn}{\downarrow}

\newcommand {\dis}{\displaystyle}

\newcommand{\beg}{\begin{equation}}
\newcommand{\en}{\end{equation}}

\newcommand{\eps}{\epsilon}

\begin{document}

\title{Nonequilibrium Cooper pairing in the nonadiabatic Regime}

\author{Emil A. Yuzbashyan$^{1}$ }

\author{Boris L. Altshuler$^{2}$}

\author{Vadim B. Kuznetsov$^3$}

\author{Victor Z. Enolskii$^4$}

\affiliation{ \phantom{a}
\vspace{0.1cm}
\centerline{$^1$Center for Materials Theory, Rutgers University,
Piscataway, New Jersey 08854, USA}
\centerline{$^2$Physics Department, Princeton University, Princeton, NJ 08544, USA}
\centerline{$^3$Department of Applied Mathematics, University of Leeds, Leeds, LS2 9JT, UK}
\centerline{$^4$Department of Mathematics, Heriot-Watt University, Edinburgh EH14 4AS, UK}
 }




\begin{abstract}

We obtain a complete solution for the mean-field dynamics of the BCS paired state with a large, but finite number of Cooper pairs in the nonadiabatic regime. We show that the problem reduces to a classical integrable Hamiltonian system and derive a complete set of its integrals of motion. The condensate exhibits irregular multi-frequency oscillations ergodically exploring the part of
the phase-space allowed by the conservation laws.  In the thermodynamic limit however the system can asymptotically reach a steady state.

\end{abstract}

\maketitle

The study of the dynamics of the BCS superconductors has a long history\cite{kopnin}. Early attempts to describe nonstationary superconductivity were based on the time-dependent Ginzburg-Landau (TDGL) equation \cite{elihu,schmid,gorkov}, which reduces the problem to the time evolution of a single collective order parameter $\Delta(t)$. The TDGL approach is valid only provided the system  quickly reaches an equilibrium with the instantaneous value of $\Delta(t)$, i.e. a local  equilibrium is established faster than the time scale of the order parameter variation, $\tau_\Delta\simeq 1/\Delta$. This requirement limits the applicability of the TDGL  to special situations where pair breaking dominates, e.g. due to a large concentration of magnetic impurities.
An alternative to TDGL  is the Boltzmann kinetic equation\cite{ber,aronov}  for the quasiparticle distribution function coupled to a self-consistent equation for $\Delta(t)$. This approach is justified only when  external parameters change slowly
on the $\tau_\Delta$ time scale, so that the system can be characterized by a quasiparticle distribution.

 Is it possible to describe theoretically the dynamics of a BCS paired state in the nonadiabatic regime when external parameters change substantially on the  $\tau_\Delta$ time scale?
In particular, an important question is whether, following a sudden perturbation, the condensate reaches a steady state on a $\tau_\Delta$ time scale or on a much longer quasiparticle energy relaxation time scale $\tau_\eps$.
 In the nonadiabatic regime   both  TDGL and the Boltzmann kinetic equations fail and one has to deal with the coupled coherent dynamics of individual Cooper pairs. Recent studies\cite{barankov,andreev,barankov2,simons} of this outstanding problem  were motivated by  experiments on fermionic pairing in cold atomic
alkali gases\cite{finally,feshbach_experiments}. The strength of pairing interactions in these systems can be fine tuned rapidly  by a magnetic field, making it easier than in metals  to access  the nonadiabatic regime experimentally.

The main result of the present paper is an explicit general solution for the {\it dynamics} of   the BCS model, which describes a spatially homogenous condensate at times
$t\ll \tau_\eps$.
We employ the usual BCS mean-field approximation, which is accurate when the number of Cooper pairs is large\cite{largeN,ander1}.    It turns out that the mean-field BCS dynamics can be formulated as a nonlinear {\it classical Hamiltonian} problem. We obtain the exact solution for all initial conditions and a complete set of integrals of motion for the mean-field BCS dynamics.

In this paper we assume that the number of Cooper pairs in the system is arbitrary large, but  {\it finite}. In this case the typical evolution
at times $t\ll\tau_\eps$ is {\it quasi}-periodic with a large number of incommensurate frequencies. The condensate exhibits irregular multi-frequency oscillations ergodically exploring the part of
the phase-space allowed by the conservation laws. The system returns arbitrarily close to its initial state at irregular time
intervals. However, the return time diverges in the thermodynamic limit for most physical initial conditions, while the solution asymptotically reaches   a {\it steady} state on
the $\tau_\Delta$ time scale.
The system thermalizes on a much larger energy relaxation
time scale $\tau_\eps$\cite{dicke}.

The dynamics of the BCS condensate following a sudden change of external parameters has
been previously discussed by a number of authors\cite{ander1,volkov,galperin,shumeiko,barankov,andreev,barankov2,simons,amin}.
Most notably, a linear analysis around the BCS ground state has been performed\cite{ander1,volkov} and  some
simple particular solutions for the nonlinear mean-field dynamics in the context of superconductivity have been  reported\cite{shumeiko,barankov}. We discuss  below how these results fit into the general picture.

We begin our description of the nonequilibrium Cooper pairing  in the  non-dissipative regime, $t\ll\tau_\eps$, with the BCS
model\cite{BCS,ander2, kaa}.
\beg
\hat H_{BCS}={\sum_{j,\sigma}\eps_{j} \hat c_{j \sigma }^\dagger \hat c_{j \sigma}-g\sum_{j, q} \hat c_{j\up}^\dagger \hat c_{j\dn}^\dagger \hat c_{ q\dn} \hat c_{ q\up}}
\label{bcs1}
\en
where $\eps_j$ are single-particle energies. The pairing is between  time reversed states $|j\up\rangle$ and $|j\dn\rangle$\cite{momentum}.
 Our goal is to determine the  evolution of  a state that was driven out of equilibrium at, say, $t=0$.

There are several equivalent ways to derive mean-field equations of motion. One can start with the BCS product state, $\prod_j\left(U_j(t)+V_j(t) c_{j\up}^\dagger \hat c_{j\dn}^\dagger\right)|0\rangle$, and use Bogoliubov-de Gennes equations for the time-dependent amplitudes $U_j(t)$ and $V_j(t)$. Alternatively, one can study the evolution of the normal, $G_j(t)=-i\langle[ \hat c_{j\up}(t),  \hat c_{j\up}^\dagger(t)]\rangle$, and anomalous, $F_j(t)=-i\langle[\hat c_{j\up}(t),  \hat c_{j\dn}(t)]\rangle$, Green's functions at coinciding times\cite{volkov}.

The most convenient for us approach to the BCS mean-field dynamics is  based on the Anderson pseudospin representation\cite{ander1}. Within this approach the mean-field equations are Hamiltonian equations of motion for a classical  spin chain.
Pseudospin-1/2 operators are related to fermion creation and annihilation operators via $\hat K_j^z=(\hat n_{j\up}+\hat n_{j\dn}-1)/2$ and $\hat K_j^-=\hat c_{j\dn} \hat c_{j\up}=(\hat K_j^+)^\dagger$.
Pseudospins are defined on empty and doubly occupied (unblocked) single-particle orbitals $\eps_j$. Singly occupied orbitals  are decoupled from the dynamics. For $n$ unblocked orbitals the Hamiltonian has the form
\beg
\hat H_{BCS}=\sum_{j=0}^{n-1} 2\eps_j \hat K_j^z-g\sum_{j, q} \hat  K_j^+ \hat  K_q^-
\label{bcs2}
\en

 The mean field approximation is accurate\cite{ander1,largeN} in the thermodynamic limit due to the infinite range of interactions between spins in the Hamiltonian \re{bcs2}.
Therefore,  the effective field seen by each pseudospin in \re{bcs2} can be replaced with its quantum mechanical average, ${\bf b}_j(t)=(-2\Delta_x(t), -2\Delta_y(t), 2\eps_j)$, where $\Delta(t)\equiv\Delta_x(t)-i\Delta_y(t)\equiv g\sum_j \langle \hat K_j^-(t)\rangle$ is the BCS gap function.
In this approximation, each spin evolves in the self-consistent field: $\dot {\hat {\bf K}}_j=i[ \hat H_{BCS}, \hat {\bf K}_j]\approx {\bf b}_j \times \hat {\bf K}_j$.
 Taking  the quantum mechanical average of these equations with respect to the time-dependent state of the system,
   we obtain for ${\bf s}_j(t)=\langle \hat {\bf K}_j(t) \rangle$
\beg
\dot {\bf s}_j={\bf b}_j \times {\bf s}_j \phantom{m} {\bf b}_j=\left(-2gJ_x, -2g J_y, 2\eps_j\right)\quad {\bf J}=\sum_{q=0}^{n-1} {\bf s}_q
\label{spins2}
\en
 The components of the classical spins $s^z_j(t)$ and $s_j^\pm=s_j^x\pm is_j^y$ are related
to Bogoliubov amplitudes and equal times Green's functions as $2s_j^z=|V_j|^2-|U_j|^2$, $s_j^-=\bar U_j V_j$ and
$G_j(t)=is_j^z(t)$, $F_j(t)=is_j^-(t)$, respectively.
Evolution equations \re{spins2} conserve the square of the average for each spin: $d{\bf s}_j^2/dt=0$. If the spins initially were in a product state, ${\bf s}_j^2=1/4$. Note also that $\Delta(t)=gJ_-(t)$.

One can check that  Eqs.~\re{spins2} are equations of motion for a {\it classical} spin Hamiltonian
\beg
H_{BCS}=\sum_{j=0}^{n-1} 2\eps_j s_j^z-g\sum_{j,q} s_j^+s_k^-
\label{class}
\en
It means that Eqs.~\re{spins2} are Hamilton equations $\dot {\bf s}_j=\{H_{BCS}, {\bf s}_j\}$ derived from Hamiltonian \re{class} using the usual angular momentum Poisson brackets
\beg
\dis \left\{ s_j^a, s_k^b\right\}=-\varepsilon_{abc} \delta_{jk} s_j^c
\label{poisson}
\en
where $a$, $b$, and $c$ stand for spatial indexes $x$, $y$, and $z$. The classical model \re{class} can be obtained from its quantum counterpart \re{bcs2}   by replacing  operators with classical dynamical variables and commutators with Poisson brackets.

  Both the classical \re{class} and quantum models \re{bcs1} and \re{bcs2} are
  integrable\cite{rich,gaudin,integr}.
  To show this, one can introduce a vector function (Lax-vector) of an auxiliary parameter $u$
\beg
{\bf L}(u)=- \frac{\hat {\bf z}}{g}+\sum_j \frac{{\bf s}_j}{u-\eps_j},
\label{lax}
\en
where $\hat{\bf z}$ is a unit vector along the $z$ axis.
Poisson brackets between components of ${\bf L}(u)$ at different values of $u$ can be evaluated using Eq.~\re{poisson}.
\beg
\left\{ L^a(v), L^b(w)\right\}=\varepsilon_{abc}\frac{L^c(v)-L^c(w)}{v-w}
\label{fundament}
\en
(Relations \re{fundament} hold for each term in \re{lax} separately; all terms Poisson-commute with each other). It follows from Eq.~\re{fundament} that the lengths of
the Lax vector  at different values of $u$ Poisson-commute:
\beg
\left\{ {\bf L}^2(v), {\bf L}^2(w)\right\}=0
\label{integgen}
\en
The scalar function ${\bf L}^2(u)$ can be represented in the form
\beg
{\bf L}^2(u)=\frac{1}{g^2}+\sum_{j=0}^{n-1} \left( \frac{2 H_j}{u-\eps_j}+\frac{{\bf s}^2_j}{(u-\eps_j)^2}\right)
\label{v}
\en
where
\beg
H_j=\sum_{k=0}^{n-1}{\lefteqn{\phantom{\sum}}}' \phantom{.}\frac{ {\bf s}_j\cdot{\bf s}_k}{\eps_j-\eps_k}-\frac{s_j^z}{g}
\label{magnets}
\en

Since Eq.~\re{integgen} holds for any $v$ and $w$, all $H_j$ Poisson-commute with each other. Therefore, each $H_j$, as well as any algebraic combination of $H_j$, defines a classical model\cite{name} that has $n$ degrees of freedom ($n$ classical spins) and $n$ integrals of motion (including itself) and thus is integrable in the usual sense\cite{arnold}. Note that the sum of $H_j$ is proportional to the $z$-component of the total spin ${\bf J}$, therefore $J_z$ is conserved by all $H_j$ and their combinations. Moreover, the following identity follows from Eqs.~(\ref{class},\ref{magnets})
\beg
H_{BCS}=-g\sum_j \eps_j H_j +\mbox{const}
\label{comb}
\en
This implies that the classical BCS model \re{class}  Poisson-commutes with all $H_j$'s and thus is also integrable. Equations~\re{magnets} and \re{comb} can be straightforwardly  quantized by replacing ${\bf s}_j\to \hat {\bf K}_j$. The resulting operators $\hat H_j$ all pairwise commute, thus showing the integrability of quantum models \re{bcs1} and \re{bcs2}.

To obtain the general solution for the mean-field dynamics of the BCS model \re{bcs1},
we follow the method of Ref.~\onlinecite{sklyanin} and introduce ${n-1}$ {\it separation} variables $u_k$ as zeros of $L_-(u)=L_x(u)-iL_y(u)$, i.e.
$\sum_j s_j^-/(u_k-\eps_j)=0$.

 Equations of motion for the variables $u_k$ are\cite{us}
\beg
 \dot u_k=2i \sqrt{Q_{2n}(u_k)}\prod_{m\ne k}(u_k-u_m)^{-1}
\label{ev}
\en
where $Q_{2n}(u)$ is the {\it spectral polynomial} defined as
\beg
Q_{2n}(u)=g^2{\bf L}^2(u) \prod_j (u-\eps_j)^2
\label{spectral}
\en
By Eq.~\re{v}, the coefficients of $Q_{2n}(u)$ depend only on the integrals of motion $H_j$.

Eqs.~\re{ev} constitute the well-known Jacobi's inversion problem solvable
 in terms of hyperelliptic theta functions\cite{theta}.
Here we outline the final answer, the details will be reported elsewhere\cite{us}.
Klenian  $\sigma$- and $\zeta$-functions of {\it genus} $G$ (in our case ${G=n-1}$)   are defined  as
\beg
\begin{array}{l}
\dis \zeta_l({\bf x})=\frac{\partial \ln\sigma({\bf x}) }{\partial x_l}\quad \sigma({\bf x})= \sum_{{\bf m}\in Z^{G} } \exp{\left[S_{\bf m}(x)/2 \right]}\\
\dis  S_{\bf m}(x)={\bf x}\cdot\eta \omega^{-1}{\bf x}+2i\pi({\bf m}\cdot\tau{\bf m}+\omega^{-1}\bf x\cdot {\bf m})\\
\end{array}
\label{sigma}
\en
where the sum is over all $G$-dimensional integer vectors ${\bf m}$, $\tau=\omega'\omega^{-1}$, and $\omega$, $\omega'$, and $\eta$ are $G\times G$ matrices of periods (see below).
The solution is
\beg
 s_j^-(t)=\langle\hat c_{j\dn}(t) \hat c_{j\up}(t)\rangle=J_-(t) r(\eps_j, t) \prod_{k\ne j}\frac{\eps_j}{\eps_j-\eps_k}
\label{trans}
\en
\beg
\Delta(t)=g J_-(t)=g\sum_j s_j^-(t) =c_n e^{-i\beta t}\frac{\sigma({\bf x}+{\bf d})}{\sigma({\bf x}-{\bf d})}
\label{Jt}
\en
Here ${\bf x}^T=i(c_1,\dots, c_{n-2}, 2t+c_{n-1})$; ${\bf d}$ is a vector of constants; $\beta=g J_z+\sum_j\eps_j$; $c_1,\dots,c_{n}$ are  constants fixed by the initial conditions, and
\beg
r(u,t)=1-\sum_{k=1}^{n-1}  (\zeta_k({\bf x}+{\bf d})- \zeta_k({\bf x}-{\bf d})+a_k) u^{k-n}
\label{R}
\en
Constants $a_k$, the matrices of periods, constant vector ${\bf d}$, are all uniquely determined\cite{victor} by the spectral polynomial $Q_{2n}(u)$, i.e. by integrals of motion.

The evolution of ${\bf s}_j(t)=\langle \hat {\bf K}_j(t) \rangle$ described by the general solution is typical of an integrable system\cite{arnold}.  It is characterized by $n$ frequencies, which in our case can be determined\cite{us} in terms of integrals of motion, and are typically incommensurate.  Note that $|\Delta(t)|$ contains only $n-1$ frequencies. The typical dynamics  is stable against perturbations destroying integrability\cite{arnold}.

Now let us discuss some  particular solutions. There are two types of equilibrium states that play an important role in the dynamics. In {\it normal} states  all spins
are parallel to the $z$-axis, $2{\bf s}_j^z=\pm1$. Since $2{\bf s}_j=\langle \hat n_j\rangle-1$, these states correspond to the ground state and excitations of the single-particle part of the Hamiltonian \re{bcs1} (Fermi gas). They are stationary within the mean-field dynamics \re{spins2}.   For a finite system, they are non-stationary for the quantum Hamiltonian \re{bcs1} and their short time dynamics is entirely driven by quantum corrections (cf. Refs.~\onlinecite{BP,vardi}).

The second type of equilibrium  states are {\it anomalous} ones, which correspond to the BCS ground state and  excitations. These states are obtained by aligning each spin in \re{class} self-consistently along the effective
magnetic field acting on it. The self-consistency condition is the BCS gap equation. As ${\bf s}_j=\langle\hat {\bf K}_j\rangle$, one can obtain the BCS wave-function and energy spectrum from anomalous equilibrium  configurations of classical spins ${\bf s}_j$.

It turns out that equilibrium states are a part of a more general scheme when the dynamics of $n$ spins degenerates to that of $m<n$ {\it collective} spins ({\it $m$-spin solutions}) governed by the same Hamiltonian \re{class} only with $m$ spins and new parameters $\mu_j$ instead of $\eps_j$. Normal and anomalous states correspond to 0- and 1-spin solutions, respectively. To construct $m$-spin solutions one has to take the Lax-vector \re{lax} to be proportional to that of a  system with $m$ spins ${\bf t}_k$, ${\bf L}(u)=[1+\sum_j b_j/(u-\eps_j) ]{\bf L}_{\bf t}(u)$, where $b_j$ are time-independent constants, and ${\bf L}_{\bf t}(u)=-{\hat {\bf z}}/g+\sum_k {\bf t}_k/(u-\mu_k)$. Then, $2(n-m)$ of $2n$, typically distinct, roots of the spectral polynomial $Q_{2n}(u)$ become doubly degenerate and $n-m$ separation variables $u_k$ are frozen in these double roots, which automatically solves the equations of motion for these $n-m$ variables. The dynamics is obtained by replacing $n\to m$ and $\eps_k\to\mu_k$ in Eqs.~(\ref{sigma}--\ref{R}) and is characterized by $m<n$ typically incommensurate frequencies.
For $m=2$ the solution is in terms of  hyperelliptic  functions of genus $G=m-1=1$, i.e. in terms of  ordinary elliptic functions.

Now let us discuss the connection of our results with the previous work.  The solutions for the mean field BCS dynamics  obtained in Ref.~\onlinecite{barankov} are 2-spin solutions in the above classification. They were used in Ref.~\onlinecite{barankov}  to describe the evolution beginning from a state
infinitesimally close to the normal ground state. In our view, the dynamics in the vicinity of this state can have additional features and deserves further analysis.

The 2-spin solutions resemble the TDGL approach in that they describe the dynamics of all pairs in terms of only two collective degrees of freedom resulting in large amplitude single frequency (periodic) oscillations of the order parameter magnitude $|\Delta(t)|$.   Mathematically, they lie on a 1d curve of   points in a multi-dimensional (infinite-dimensional in the thermodynamic limit) space of possible values of integrals of motion.
 The situation with other few spin solutions is similar\cite{asymp}.
In contrast, the general solution we obtained here  {\it typically} has a large (infinite in the thermodynamic limit) number of incommensurate frequencies and a substantially reduced amplitude. The difference between the general and few spin solutions is clear in a linear analysis\cite{ander1,volkov} around the BCS ground state that displays normal modes with frequencies $\omega_k=2 \sqrt{\eps_k^2+\Delta_0^2}$, where $\eps_k$ are single-particle energies and $\Delta_0$ is the equilibrium  order parameter. In the linear regime, the general solution becomes an arbitrary superposition of all normal modes, while few spin solutions single out all, but few modes. For example, 2-spin solutions of Ref.~\onlinecite{barankov}  correspond to a single normal mode with
a frequency $2\Delta_0$.

In conclusion, we have obtained the explicit general solution for the mean-field dynamics of the BCS paired state and discussed a number of special cases including two types of equilibrium states and few spin solutions. A still open problem is to fully analyze the solution in the thermodynamic limit.    It is also desirable to better understand the dynamics in the vicinity of normal  states where  quantum effects become important. Finally, it is interesting to identify experimental setups where peculiar features of the nonequilibrium Cooper pairing in the nonadiabatic regime can be observed  such as e.g. cold Fermi gases.

We are grateful to I. Aleiner, V. Falko, L. Glazman, L. Levitov, A. Millis,  A. Polyakov, and O. Tsyplyatyev for stimulating discussions. This research was supported by NSF DMR 0210575 and by ARO/ARDA
(DAAD19-02-1-0039).

{\it Note Added.} Recently, we became aware of a publication\cite{warner} that is in agreement with some of our
conclusions -- that the initial dynamics of the normal ground state is driven by quantum corrections and that  the
system can reach a steady state at large times.


\begin{thebibliography}{99}
\bibitem{name} Each of $H_j$ can be viewed as a Hamiltonian of a system of classical spins, these Hamiltonians are known as classical Gaudin magnets\cite{gaudin}.
\end{thebibliography}
\end{document}